\documentclass[aps,twocolumn,superscriptaddress,pra,10pt,floatfix]{revtex4-2}
\usepackage{graphicx}
\usepackage{dcolumn}
\usepackage{bm }
\usepackage{paralist}
\usepackage{braket}
\usepackage{hyperref}
\usepackage{natbib}
\usepackage{xcolor}
\usepackage{upgreek}
\usepackage{comment}

\newcommand{\SiN}{Si$_3$N$_4$ }
\newcommand{\LiN}{LiNbO$_3$ }

\begin{document}

\title{Technologies for Modulation of Visible Light and their Applications }
\author{Sanghyo Park}
\affiliation{Department of Physics, Korea Advanced Institute of Science and Technology, Daejeon 34141, South Korea}
\author{Milica Notaros}
\affiliation{Research Laboratory of Electronics, Massachusetts Institute of Technology, Cambridge, Massachusetts, 02139, USA}

\author{Aseema Mohanty}
\affiliation{Department of Electrical \& Computer Engineering, Tufts University, Medford, Massachusetts 02155, USA}

\author{Donggyu Kim}
\affiliation{Department of Physics, Korea Advanced Institute of Science and Technology, Daejeon 34141, South Korea}
\author{Jelena Notaros}
\affiliation{Research Laboratory of Electronics, Massachusetts Institute of Technology, Cambridge, Massachusetts, 02139, USA}
\author{Sara Mouradian}
\affiliation{Department of Electrical \& Computer Engineering, University of Washington, Seattle, Washington 98195, USA}

\begin{abstract}

Control over the amplitude, phase, and spatial distribution of visible-spectrum light underlies many technologies, but commercial solutions remain bulky, require high control power, and are often too slow. Active integrated photonics for visible light promises a solution, especially with   recent materials and fabrication advances. In this review, we discuss three growing application spaces which rely on control of visible light: control and measurement of atomic quantum technologies, augmented-reality displays, and measurement and control of biological systems. We then review the commercial dynamic surfaces and bulk systems which currently provide visible-light modulation and the current state-of-the-art integrated solutions. Throughout the review we focus on speed, control power, size, optical bandwidth, and technological maturity when comparing technologies. 
    
\end{abstract}

\maketitle

Control over the temporal and spatial characteristics of light has revolutionized fields from communication to the lithography tools which enable the semiconductor industry. In the 19th century, Pockels and others investigated how external electric and strain fields could change the optical properties of materials, enabling a host of bulk optical elements for manipulating light. These bulk modulators are the current workhorse for light modulation with near-unity efficiencies and modulation speeds up to GHz. However, these components provide limited spatial control, are generally large, and require high drive power. 

Since the early 1990s, miniaturized and integrated systems have been recognized as the future of active light control~\cite{koch_1991_photonicsperspective} and integrated photonics and controllable surfaces promise to reduce both the size and power requirements of visible-light modulation devices. Indeed, for light with frequencies below the bandgap of Silicon (Si) ($< 1.12$\,eV, $\lambda > 1100$\,nm) great progress has been made in developing compact, fast modulators based on integrated photonics~\cite{Siew2021Mar,rahim_2018_Si,rickman_2014_siphotonics}. The trillion-dollar electronic chip industry relies on complementary metal-oxide semiconductor (CMOS) fabrication, and by leveraging this industry, Si integrated photonics has been able to move from proof-of-principle demonstrations to deployment in data centers and high-rate intercontinental communication~\cite{Margalit2021May}. However, the miniaturization and commercialization of devices to control visible light has so far lagged behind. 

Control of visible light at low per-channel drive power, compact form-factor, and high frequencies will enable the growth of a wide variety of technologies. In particular, control and measurement of quantum systems, augmented-reality displays, and imaging and control of biological systems are three application areas that have enjoyed significant growth in the last decade, but which are hampered by the capabilities of current commercially-available visible-light modulators. This explosion of application areas coupled with materials and processing developments has accelerated the research in and development of technologies for visible-light modulation. Here, we review the needs of these three application spaces and the current state-of-the-art for visible-light modulation technologies. We aim to provide a review  of the application space to motivate and focus further research in scalable visible light modulation platforms and a review of the optical state-of-the-art for researchers in application fields who are interested in scaling up modulation of visible light.

In Sec.~\ref{Sec:Applications}, we discuss the needs of three application spaces in detail - quantum control and readout of atomic qubits, augmented-reality displays, and control of biological systems. These technologies serve as an anchor point for the figures of merit of the various visible-light modulation platforms discussed in the following sections. In Sec.~\ref{Sec:Surfaces}, we discuss reflective and transmissive controllable surfaces. These, mainly commercial, solutions enable current application systems but suffer from high per-channel power and size, and often cannot reach the necessary modulation speeds. In Sec.~\ref{Sec:Integrated}, we discuss integrated solutions to visible-light modulation. These nascent technologies need significant progress to reach the maturity needed for systems-level integration, but hold great promise.  Finally, we conclude with a perspective on future research directions needed to transition these technologies out of the laboratory and into the field. 

\section{Applications}
\label{Sec:Applications}
Many fields would benefit from many-channel, high-speed, low-power modulation of visible light. Here, we focus on three applications: (1) quantum control of atomic qubits; (2) augmented-reality displays; and (3) biological applications, such as optogenetics and microscopy. We discuss the performance requirements and the current state-of-the-art for each of these applications. 

Additionally, there are a plethora of other potential applications which may be enabled by improved modulation of visible light. For instance, both classical and quantum optical computing would benefit from moving to visible wavelengths to  reduce component size. Similarly, most light detection and ranging (LiDAR) demonstrations use telecommunications light, but visible LiDAR can outperform infrared systems in underwater  applications due to reduced absorption~\cite{wu_2017_bathcomms,notaros23ol,desantis23ipc}. Finally, visible light communication or Li-Fi~\cite{Matheus2019Apr} has applications in indoor, high-accuracy positioning or when electromagnetic interference must be avoided. The wavelengths of interest for all these applications are shown in Fig.~\ref{Fig:Applications}.

\subsection{Quantum Technologies}

Quantum technologies promise improvements over their classical counterparts in communication~\cite{Gisin2007Mar}, sensing~\cite{Degen2017Jul}, timekeeping~\cite{Mehlstaubler2018Apr,Ludlow2015Jun}, and computation~\cite{Nielsen2010Dec}. However, to provide utility beyond proof-of-principle demonstrations it will be necessary to reduce the size, weight, and power of all components while simultaneously improving stability and capability. Neutral atom and charged ion qubits are two of the leading platforms for quantum information processing (QIP) and both rely on ultraviolet, visible, and near infrared light for control and measurement. Although we do not discuss them explicitly here, quantum control of molecules~\cite{Mitra2022Apr} and a broad range of solid state atomic-like defects, for example in diamond~\cite{Bradac2019Dec,Pezzagna2021Mar} or SiC~\cite{Castelletto2020Mar}, also rely on visible light modulation, with requirements similar to neutral atoms and trapped ions.

\begin{figure}[htbp]
\begin{center}
\vspace{0.5cm}
\includegraphics[width=0.48\textwidth]{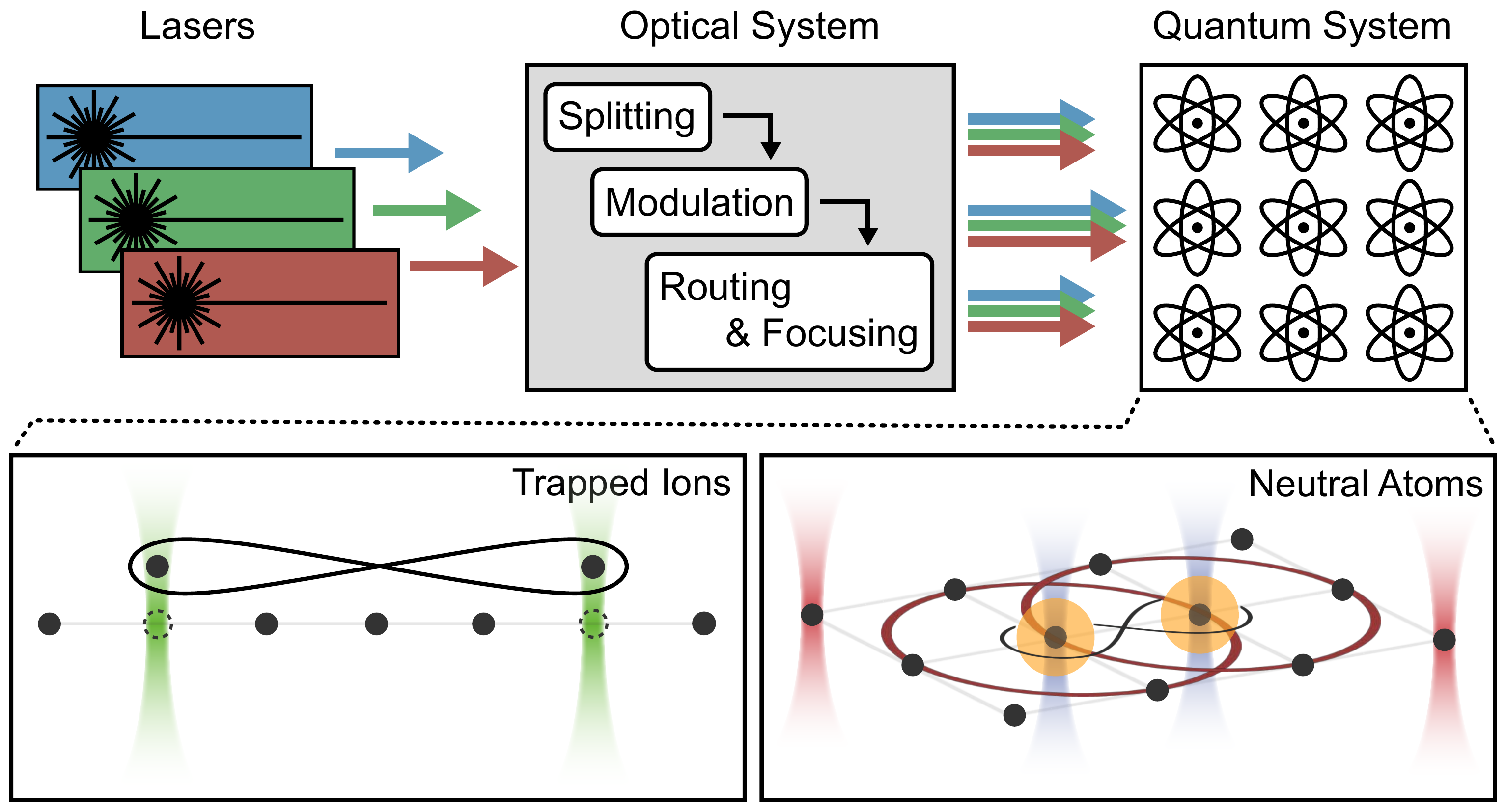}
\caption{\small Block diagram depicting the optical control system for quantum technologies. Laser light is routed through an optical system which splits each source into many channels, modulates each channel, and shapes the light to control the quantum system.}
\label{Fig:QBlock}
\end{center}
\end{figure}

The majority of experimental demonstrations of atomic QIP have relied on bulk crystals to modulate the light needed to control and readout the quantum state of the system. However, to scale these systems to the millions of qubits necessary for fault-tolerant quantum computation~\cite{Beverland2022Nov}, it will be necessary to actively modulate the amplitude, frequency, phase, and often position of millions of channels simultaneously at wavelengths across the UV and visible spectrum. This is untenable with current solutions. Similarly, deployable quantum sensors, clocks, and quantum network nodes will need compact, stable optical control. It is thus clear that integrated photonics will play a critical role in future quantum technologies~\cite{Luo2023Jul,Pelucchi2022Mar,Zhang2023Aug}.

\subsubsection{Trapped Ions}
\begin{figure*}[htbp]
\begin{center}
\vspace{0.5cm}
\includegraphics[width=0.98\textwidth]{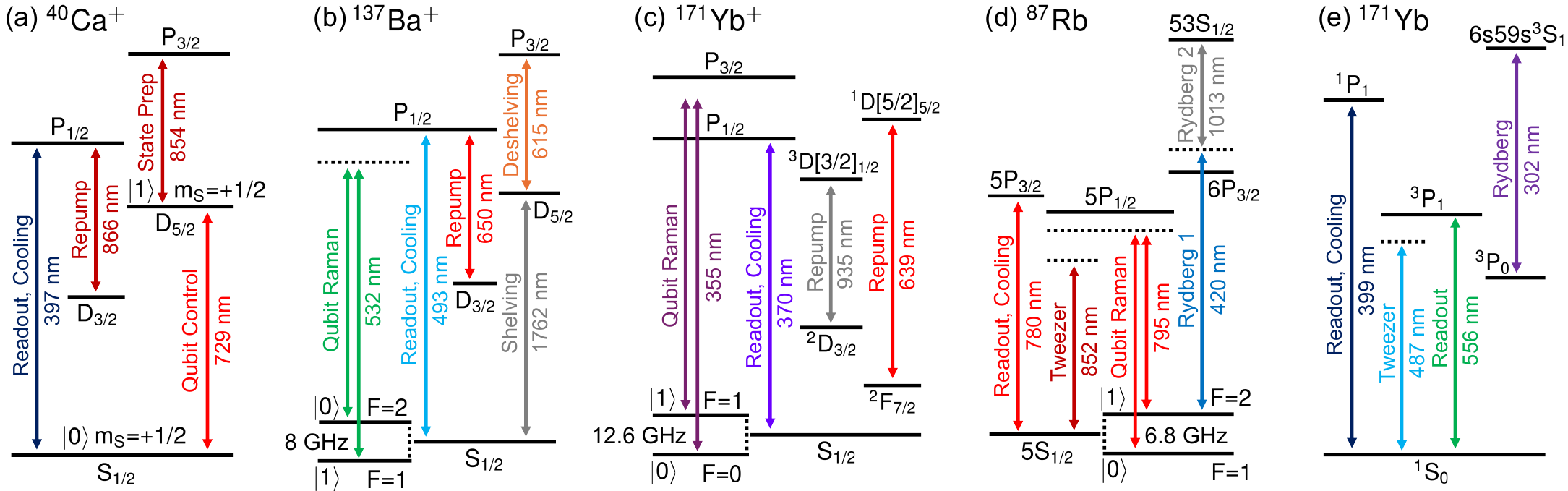}
\caption{\small Electronic level structure for $^{40}$Ca$^+$, $^{137}$Ba$^+$, and $^{171}$Yb$^+$ as three representative ions, and for $^{87}$Rb and $^{171}$Yb as two representative neutral atoms. Each transition is labeled with its function for quantum control. Light is scattered off \emph{Readout} transitions to measure the quantum state. \emph{Cooling} transitions are used to reduce the motion of a trapped atomic qubit to near the ground state. \emph{Shelving} transitions are used to hide some atomic population in along lived state, usually during quantum state discrimination. \emph{Repump} transitions are used to remove population from a state where it may otherwise be trapped. \emph{Rydberg} transitions are used to excite an atom to a highly excited state. \emph{Tweezer} beams are detuned from a dipole transition and are used to confine an atom. Finally, \emph{Qubit} transitions are used to control the state of the qubit.}
\label{Fig:IonLevels}
\end{center}
\end{figure*}
Trapped ions are a promising platform for quantum simulation~\cite{Blatt2012Apr,Zhang2017Nov}, quantum sensing~\cite{Kotler2011May,Gilmore2021Aug,Mouradian2021Mar}, optical clocks~\cite{Brewer2019Jul}, and gate-based quantum computation~\cite{Bruzewicz2019Jun,Ryan-Anderson2022Aug,Chen2023Aug}. Indeed, the state-of-the-art systems provide the highest reported quantum volume to date~\cite{BibEntry2024Feb,Moses2023Dec}. In trapped-ion quantum technologies, atomic ions (generally singly charged) are trapped in ultrahigh vacuum using electromagnetic fields. Quantum information is stored in the coherence between two stable states of each ion.  Each ion qubit is by nature identical and can be well-isolated from the environment. Furthermore, ions trapped together in the same confining potential share motional modes that can be used to directly mediate entanglement between qubits~\cite{Sorensen1999Mar}, leading to natural all-to-all connectivity within a chain of ions trapped in a single potential. Entanglement between ions in different potentials can be mediated by either physically shuttling the ions between trapping regions~\cite{Moses2023Dec} or optically~\cite{Moehring2007Sep}.


Fig.~\ref{Fig:IonLevels} shows relevant transitions for controlling (a) $^{40}$Ca$^+$ and (b) $^{137}$Ba$^+$, and (c) $^{171}$Yb$^+$, three commonly used ions. Other ions and isotopes exhibit similar level structures, though with different transition frequencies. As evidenced in Fig.~\ref{Fig:IonLevels}, control of even a single species can span the spectrum from the ultraviolet to the near infrared. Additionally, it is likely that a final trapped-ion QIP architecture will employ multiple ion species~\cite{Moses2023Dec} or encodings~\cite{Allcock2021Nov} for mid-circuit cooling and measurement.

Trapped-ion qubits can generally be separated into two categories -- optical and ground-state qubits. For example, $^{40}$Ca$^+$ optical qubits are encoded between one state in the $4^2$S$_{1/2}$ manifold and one state in the metastable $3^2$D$_{5/2}$ manifold~\cite{Haffner2008Dec}. This is a weakly-allowed quadrupole transition with an optical transition at 729\,nm. Driving single qubit gates requires only modest power ($\sim10\mu$W per qubit) but stringent frequency and phase stability~\cite{Bruzewicz2019Jun}. Conversely, in $^{137}$Ba$^+$~\cite{Dietrich2009Mar} and  $^{171}$Yb$^+$~\cite{Nop2021Dec}, qubits are generally encoded in the S$_{1/2}$ hyperfine levels. These qubits have GHz-scale splittings and can be driven directly with microwave fields or optically through two-photon Raman transitions~\cite{Gaebler2016Aug}. Raman control requires high power ($\sim10$\,mW per qubit) and often low wavelengths ($<$ 400\,nm), but the requirements for frequency and phase stability are relaxed as only the difference frequency and phase between the two Raman beams must be precisely controlled. For both qubit encodings, single qubit gate times are around 1\,$\mu$s, and two qubit gate times are around 100\,$\mu$s~\cite{Bruzewicz2019Jun}, necessitating amplitude, phase, and frequency control at speeds above 1\,MHz. 

For all trapped-ion qubits, cooling~\cite{Eschner2003May}, state preparation~\cite{Haffner2008Dec}, and readout~\cite{Myerson2008May} are performed with optically accessible dipole transitions. These operations generally also require MHz switching speeds and around 10$\mu$W of power per beam.

Currently, the majority of trapped ion experiments rely on bulk acoustic optic modulators (Section~\ref{Sec:AOM}) for light modulation due to their commercial availability and ease of use. Some initial work has been done to scale the optical control for trapped-ion quantum technologies, both for computing and timekeeping. Passive waveguides and miniturized focusing optics can route and focus light to multiple trapping sites in parallel~\cite{Day2021Feb,Binai-Motlagh2023Jul}. Optical routing with waveguides integrated with trapping electrodes has been demonstrated for improved vibration compensation~\cite{Mehta2016Dec}, multi-wavelength control~\cite{Niffenegger2020Oct}, cooling~\cite{Hattori2022Oct,hattori24ol,sneh22fio,corsetti23fio}, multi-site control~\cite{Mordini2024Jan}, and single qubit~\cite{Vasquez2023Mar} and two qubit~\cite{Mehta2020Oct} gates. Beam shaping and focusing optics can likewise be tailored to trapped ion applications~\cite{Beck2023Jun}, including individual ion addressing~\cite{Shirao2022Jun,Hu2022Dec,Shih2021Apr} and improved photon collection efficiency~\cite{Streed2011Jan,Ghadimi2017Jan}. 

Recently, piezo-actuated waveguide modulators (see Sec.~\ref{Sec:IntStrain}) were used to control the state of a $^{40}$Ca$^+$ trapped-ion qubit with MHz switching speeds and nearly 40\,dB on-off extinction~\cite{Hogle2023Jul}. With the integrated modulators they measured 99.7\,\% single qubit gate fidelity with randomized benchmarking and performance that nearly matched the performance of standard acoustic optic modulation. However, they found that the integrated modulators were significantly less stable, highlighting that significant device and systems-level improvements are needed to reach the stringent requirements needed for quantum control.  

\subsubsection{Neutral Atoms in Optical Tweezer Arrays}
Neutral-atom arrays are an another leading platform for quantum metrology~\cite{Madjarov2019Dec,Young2020Dec,Norcia2019Oct,Shaw2024Feb}, simulation~\cite{Bernien2017Nov,Scholl2021Jul,doi:10.1126/science.abi8794,Ebadi2021Jul,Bluvstein2021Mar,Chen2023Apr,Scholl2021Jul}, and information processing~\cite{Bluvstein2022Apr,Graham2022Apr,Ebadi2022May,ak_patent_2022,Bluvstein2024Feb}. Recently, they have also enabled impressive advances towards fault-tolerant quantum processing~\cite{Bluvstein2024Feb}. The neutral atoms are laser-cooled and then loaded into programmable arrays of optical tweezers~\cite{Kim_2019}. Defect-free arrays with hundreds to thousands of atoms are possible~\cite{Ebadi2021Jul,Schymik2021Sep,Norcia2024Feb,Manetsch2024Mar,Pause2024Feb}. The neutral-atom qubits are two stable internal states of the atoms with a long coherence time. Like ion qubits, they are identical and well-isolated from their environment. 

In neutral atom QIP, lasers are used to trap and control the position of individual neutral-atom qubits and mediate single and multi-qubit gates. For entangling gates, laser beams selectively excite target atoms to highly excited Rydberg states, enabling mutual interactions via the Rydberg blockade effect~\cite{Urban2009Feb,Evered2023Oct}. Nearby atoms cannot be simultaneously promoted to the Rydberg state, naturally implementing the conditional-NOT~\cite{Maller2015Aug,Graham2022Apr,Bluvstein2022Apr,Evered2023Oct} and Toffoli~\cite{PhysRevLett.123.170503} gates between the qubits. This effect can also be used to control the system's many-body Hamiltonian for quantum simulation and optimization~\cite{Omran2019Aug,doi:10.1126/science.abi8794,Ebadi2021Jul,Ebadi2022May}. Moreover, the dynamic programming of tweezer positions allows the entangled qubits to be transported across an array. This provides reconfigurable, all-to-all connectivity essential for efficiently implementing hard problems~\cite{Ebadi2022May, Kim2022Jul}, quantum error-correction codes~\cite{Bluvstein2022Apr,xu2023constantoverhead} and fault-tolerant information processing~\cite{Scholl2023Oct,Wu2022Aug,Bluvstein2024Feb}. 

Control of neutral-atom qubits uses multiple colors of laser light (Fig.~\ref{Fig:IonLevels}). For instance, an array of green laser foci creates dipole trap arrays for $^{171}$Yb qubits~\cite{Saskin2019Apr}. Driving the ground-to-Rydberg transition of Rubidium uses 297 nm (the combination of 480 and 780 nm or 420 and 1013 nm) lasers for the single- (two-) photon transition~\cite{Li2019Apr_Atom,Chew2022Oct, Omran2019Aug}. Laser cooling and trapping neutral atoms requires kHz timing control, and the timescales for quantum gates typically ranges from 1 MHz~\cite{Bernien2017Nov, Chen2023Apr} to 100 MHz~\cite{Evered2023Oct,Chew2022Oct,Graham2019Dec} depending on the implementation. Moreover, since the atom arrangement determines the qubit connectivity~\cite{Bluvstein2022Apr,Ebadi2022May,Kim2022Jul} the dynamic control of spatial modes combined with the fast temporal modulation (e.g., controlling the position of the lasers providing the trapping potentials) is essential for assembling and rearranging atom arrays and implementing local qubit control~\cite{Bluvstein2024Feb}.

The required control currently utilizes different modulation mechanisms. For optical tweezers, liquid-crystal-on-silicon (LCoS) spatial light modulators (Section~\ref{Sec:IntLC}), mechanical mirror arrays (Section~\ref{Sec:Mech}), and acousto-optic modulators (Section~\ref{Sec:AOM}) are used to generate and rearrange an array of laser beams from a single incident laser beam. High-speed optical modulators such as acousto-optic and electro-optic modulators are used for resonant control of the atoms. Combinations of such modulation mechanisms also play an essential role to enlarge the control capabilities. However, development of advanced photonic systems will further extend the range of quantum circuits and simulations implemented on the neutral-atom array systems.\\



To summarize, quantum control applications require high speed ($>$ MHz) modulation for wavelengths across the UV and visible spectrum. Size, optical loss, and control power will be important for deployable systems and will be important to scale to the millions of channels needed for utility-scale quantum information processing.  Although not discussed in detail in this review, optical power handling will also be a particular concern, especially for two-photon control as necessary for Raman single and multi-qubit gates in trapped ions and optical tweezer generation and multi-photon excitation for neutral atoms. Additionally, isolation between on and off states is an important consideration for all of the quantum operations discussed here, with eventual goals of isolation better than 40\,dB to reach the gate errors needed for fault-tolerant quantum computation. Both power handling and isolation are figures of merit that are not often optimized for in current integrated visible-light solutions but should beconsidered moving forward. The importance of the main figures of merit discussed in this review (speed, control power, size, optical bandwidth, and technical maturity) are qualitatively depicted in Fig.~\ref{Fig:Applications}(a). 

\subsection{Augmented-Reality Displays} 

Augmented-reality display technology is another emerging application area where visible-light modulators are essential. In many situations, such as military operations and medical procedures, access to real-time information can be a key determinant for success~\cite{livingston2011military,sielhorst2008advanced,khor2016augmented}. Traditionally, this information has been displayed using head-down or head-up displays~\cite{liu2004comparison}. However, there have recently been extensive efforts to develop head-mounted displays capable of relaying information directly in the user's field of view (FOV)~\cite{xiong2021augmented,kress2013review}. These head-mounted displays enable the user to remain engaged with their surroundings while referencing information to real-world objects and events for an augmented-reality experience.

Typical commercially-available augmented-reality head-mounted displays employ an optical relay system for each eye, wherein an image produced by a microdisplay is magnified using a system of lenses to generate an image superimposed on the external scene at a single virtual focal plane in the user's FOV~\cite{xiong2021augmented,kress2013review}. However, the bulk-optics components utilized in these typical head-mounted displays result in large, heavy, and indiscreet systems. Additionally, head-mounted displays typically employ low-luminance microdisplays (approximately 1000 cd/m$^2$), which render the systems inadequate for use in ambient daylight conditions (though recent developments in high-density microLEDs show promise for improved luminance \cite{shin2023vertical}). Moreover, typical head-mounted displays utilize optical relay systems with limited FOVs (limited to ${<}\,40^\circ$ compared to the 60$^\circ$ near-peripheral FOV of the human eye). Finally, typical head-mounted displays magnify the microdisplay image such that it appears at a single virtual focal plane and are not capable of producing holographic images with full depth cues. This lack of depth information results in users experiencing eyestrain and headaches that limit long-term and wide-spread use of these displays, an effect known as the vergence-accommodation conflict~\cite{kramida2015resolving,hoffman2008vergence,shibata2011visual}. Given these limitations, there is a growing need for a discrete, mobile, large-FOV, and high-brightness augmented-reality head-mounted display with full binocular and monocular depth cues.

The field of visible-light integrated photonics has the potential to enable solutions that address this need. There have been a number of recent proposals and initial passive demonstrations of integrated-photonics-based near-eye displays that utilize holographic image projection to emit full phase fronts and resolve the vergence-accommodation conflict~\cite{notaros19cleoPassive,raval2018integrated,martinez2018see,meynard2023lpcvd}. As these systems progress to demonstrations with dynamic holographic video display functionality, they will require advances in integrated visible-light modulators. Specifically, as summarized in Fig~\ref{Fig:Applications}(a), these integrated-photonics-based augmented-reality displays will require integrated modulators with visible-spectrum-spanning functionality for multi-wavelength RGB operation and with compact sizes and low-power operation for densely-integrated high-resolution systems. However, they will not require high modulation speeds since the refresh rate of the human eye is only approximately 60 Hz.


\subsection{Biology} 
\begin{figure}[htbp]
\begin{center}
\vspace{0.5cm}
\includegraphics[width=0.48\textwidth]{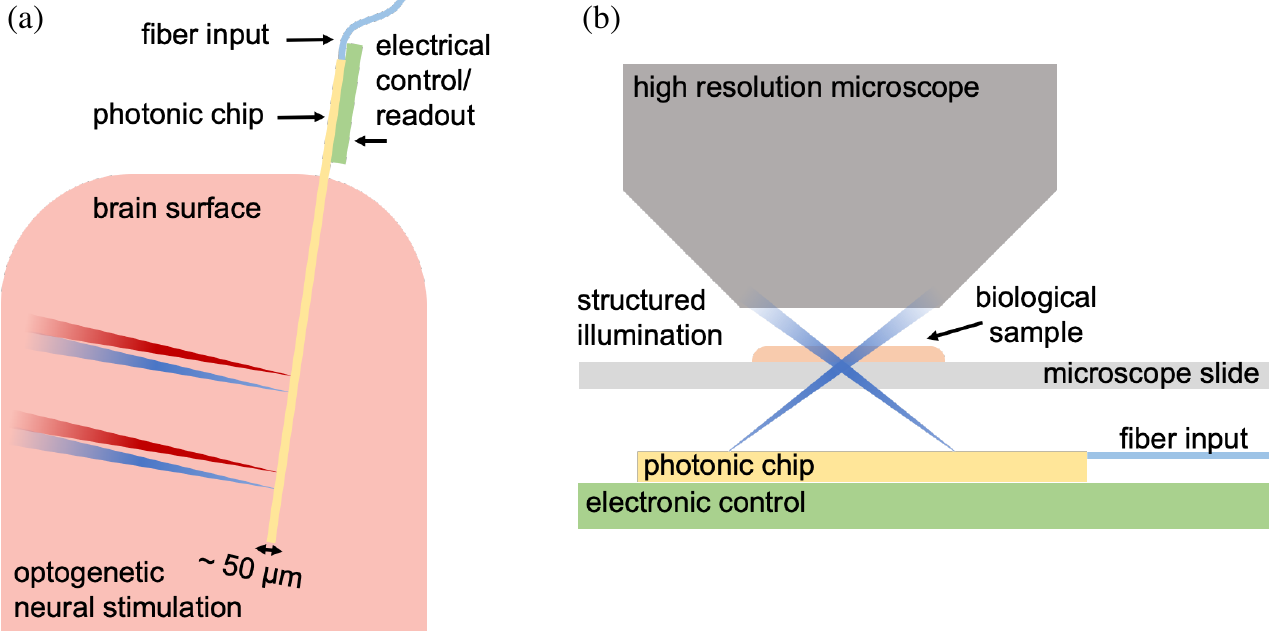}
\caption{\small a) Implantable optogenetic neural probe based on a photonic chip for control of neural activity through multicolor excitation. b) photonic chip based far-field structured illumination for high resolution imaging.}
\label{Fig:BioApp}
\end{center}
\end{figure}

Visible wavelength optical systems play a critical role in the spatiotemporal control and readout of cells and their dynamics within complex biological systems. Light can be used to target optical  transitions of molecules and proteins within cells to trigger cell functions or read out cell states~\cite{emiliani_optogenetics_2022}. These techniques typically rely on microscopes and table-top optical setups including liquid crystal spatial light modulators, galvanometer mirrors, digital micromirrors (Section~\ref{Sec:Mech}), and acousto-optic deflectors (Section~\ref{Sec:AOM}) to generate and reconfigure three-dimensional optical patterns. Due to the size, complexity, and speed of traditional free-space optical systems, biological experiments are often restricted to isolated tissues or cells or in-vivo studies of immobilized animals. High-speed visible photonic integrated circuits offer a path towards miniaturization and scalability of these optical systems enabling portable, wearable, and implantable  devices. Here we outline two biological applications that require high-speed, low power, and scalable visible modulation: optogenetic neural stimulation and high resolution microscopy. Schematics of these two applications using active intgrated photonic circuits for light delivery is shown in Fig.~\ref{Fig:BioApp}.

\subsubsection{Optogenetic Neural Stimulation}
The brain is one of the most complex and dynamic organs, containing more than a billion individual neurons and a trillion connections between them. Optogenetics has enabled the study of neural circuits with unprecedented spatiotemporal resolution and cell-specificity. Optogenetic actuators and fluorescent reporters are introduced into cells to allow specific wavelengths of light to control and readout neural activity of genetically-defined cell-types in particular brain regions, enabling the exploration of neural circuits. Controlled perturbation experiments can now be performed to elucidate the neural circuits involved in neurological and mental health illnesses and brain-machine interface technologies.

These actuators and reporters rely on optical transitions that lie in the visible wavelength range from 450\,nm to 600\,nm~\cite{emiliani_optogenetics_2022} as shown in Fig.~\ref{Fig:Applications}. For neural activity control, an opsin is used to render a neuron light-sensitive for optogenetic stimulation: when a neuron is exposed to light within the opsin's excitation bandwidth, an action potential is either excited or inhibited. Channelrhodopsin-2 (ChR2), a common opsin, is typically switched between 1-10\,Hz, while ChETA, a high-speed opsin, can switch up to 200\,Hz. Although the speed requirement for individual neuron control is slow, high-speed excitation (i.e. MHz or GHz) could enable efficient, large-volume excitation. The amount of optical power necessary for one photon optogenetic excitation is approximately 5-10\,$\mu$W per cell. 

For neural activity readout, fluorescent reporters such as genetically-encoded calcium and voltage indicators (GECIs and GEVIs)  are used to produce a fluorescence signal during an action potential either indirectly through the increased presence of calcium from ion channels or directly through a change in the membrane voltage. The temporal resolution of GECIs is greater than 50\,ms and a GEVIs can have sub-ms dynamics~\cite{gong_high-speed_2015,knopfel_optical_2019}. Although the GEVIs have faster dynamics which can more accurately resolve single neural spikes, there is ongoing research into improving their efficiency with implications for ultimate imaging depths and volumes.

The state-of-the-art optogenetic neural control and readout technologies rely on two-photon holography for simultaneous 3D optical excitation and calcium imaging using table-top optical microscopy systems~\cite{adesnik_probing_2021,mardinly_precise_2018,papagiakoumou_scanless_2020,shemesh_temporally_2017,zhang_closed-loop_2018,yang_simultaneous_2018}. This has led to impressive demonstrations of all-optical closed-loop neural control during behavior~\cite{zhang_closed-loop_2018}. However, multiphoton techniques cannot reach deeper than 0.5-1\,mm due to tissue scattering within the brain. Implantable devices based on MicroLEDs and nanostructured optical fibers have been used to push beyond this depth limit enabling demonstrations within freely-moving and behaving animals~\cite{spagnolo_tapered_2022,pisanello_dynamic_2017,ko_optogenetic_2022,voroslakos_hectostar_2022,kim_injectable_2013}. However the scale and complexity of optical pattern generation has been limited by either power consumption constraints or optical resolution.

To push beyond tissue scattering depth limits, photonic integrated circuits have been incorporated into minimally-invasive implantable probes with cellular-scale dimensions, approximately 20-100\,$\mu$m. Neural probes have been predominantly based on passive photonic integrated circuits that use off-chip reconfiguration mechanisms such as wavelength tuning or multicore fibers~\cite{segev_patterned_2016,shim_multisite_2016,sacher_optical_2022,lanzio_small_2021,reddy_low-loss_2018}. 

Active photonic integrated circuits with integrated electrical recording have utilized a photonic switching network for single-cell, sub-millisecond neural stimulation and recording in deep-brain regions previously inaccessible by free-space optics~\cite{mohanty_reconfigurable_2020}. The switching network was based on thermo-optic modulation of silicon nitride waveguides allowing for 200\,Hz neural switching at the temporal limit of the ChETA opsin. Deterministic multi-neuron spike generation was demonstrated with high spatiotemporal resolution. Thermal issues were avoided through control of the duty cycle and pulse frequency. Recent advances in monolithic integration of photonics and electronics have scaled up the number of optical emitters and recording electrodes~\cite{chen_implantable_2022,neutens_dual-wavelength_2023}. For a comprehensive review of chip-scale all-optical control and readout methods, please refer to the following perspective by Moreaux~\cite{moreaux_2020_optogenetics}. 

The scaling of active implantable photonic devices is ultimately limited by power consumption, drive voltages, and speed of the modulators which, in turn, control the complexity of optical patterns that can be produced. The optical pattern complexity directly impacts the number of single neurons or neuron populations that can be addressed. Although biological processes generally occur at millisecond timescales (Hz), the faster speeds (MHz-GHz) allow for addressing larger populations and real-time, closed-loop interrogation of the biological systems as biological processes and behaviors are complex and dynamic. It will be necessary to limit power consumption and drive voltages of active devices in proximity to biological systems to prevent disruption to natural processes or obfuscation of small neural signals. Although these circuits may remain outside of the brain, depending on dimensions of the probe and heat sinking methods, residual increases in temperatures along the shank can occur as the number of modulators increases. An ultimate limit is imposed by the brain temperature which cannot be raised more than 1-2$^{\circ}$\,C. This demands that much less than a mW of power is dissipated per modulator for a truly large-scale system. The future potential for wireless or battery operation of photonic devices will further restrict the power consumption and drive voltages needed for reconfiguration. 

\subsubsection{High-resolution Microscopy}
High-resolution microscopy techniques rely on 3D patterning and structuring of light to overcome limitations in resolution, FOV, and photo-damage during biological imaging, revealing sub-cellular structure and dynamics. Fluorescence microscopy uses fluorescent-labeled proteins or the endogenous label-free fluorescence within a sample. These can span wavelengths from UV through the visible spectrum. For example, light-sheet and structured illumination fluorescence microscopy both rely on shaped excitation to improve performance. In light-sheet microscopy, a Gaussian or Bessel beam is formed orthogonal to the detection path to provide a sheet of localized fluorescence signal across a wide FOV, leading to faster, larger volumetric imaging with high axial resolution and reduced photodamage by eliminating out-of-focus signal~\cite{stelzer_light_2021,glaser_multi-immersion_2019,chen_lattice_2014}. In structured illumination microscopy, sinusoidal excitation patterns are used to form translated or rotated Moire fringes in which high frequency information within the sample can be revealed leading to super-resolution imaging~\cite{gustafsson_surpassing_2000}. However, the typical table-top apparatus to generate and scan the excitation light is bulky, limiting the volumes and types of samples that can be studied. The miniaturization of these techniques could provide a path for widely accessible high-resolution microscopy, eventually used for diagnostics, point-of-care settings, or even wearable devices for continuous monitoring. 

Passive photonic integrated circuits have demonstrated a path towards miniaturization of these microscopy techniques. An implantable light-sheet imaging probe generating multiple light-sheets at previously inaccessible depths was demonstrated using passive reconfiguration of an array of grating emitters using a multi-core fiber~\cite{sacher_implantable_2021}. A chip-based solution used interference within waveguides to form the needed sinusoidal pattern across a larger area and with smaller minimum features. This improved the resolution and the FOV while also significantly reducing the footprint of the microscope~\cite{helle_structured_2020,diekmann_chip-based_2017}. As in the free-space domain, these techniques typically require scanning and reconfiguration to increase the imaging volume and speeds.

Recently, thermally-controlled aluminum oxide photonic integrated circuits have been used for far-field UV label-free structured illumination microscopy~\cite{lin_uv_2022,lin_uv_2023}. The phase-shifters are used to precisely control the phases of multiple beams for high contrast interference above the chip. Super-resolution is achieved using a standard wide-field microscope in conjunction with chip illumination. Although the power constraints for imaging may not be as restrictive as implantable devices, similar power and voltage drive constraints remain to prevent disruption of biological processes. The reconfiguration speed of the excitation light will impact imaging frame rates and volumes, so high-speed visible photonic integrated circuits that push beyond the speeds of state-of-the-art acousto-optic deflectors and spatial light modulators remains a critical goal.

\subsection{Summary}
\begin{figure}[htbp]
\begin{center}
\vspace{0.5cm}
\includegraphics[width=0.48\textwidth]{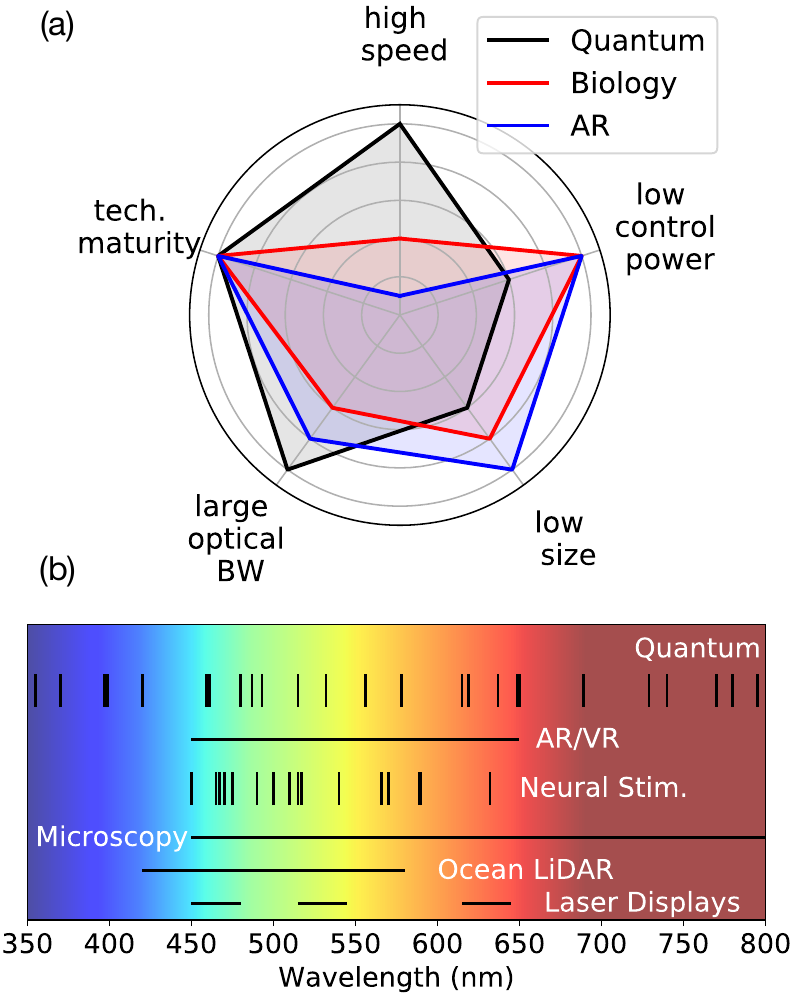}
\caption{\small (a) Qualitative depiction of the importance of each of five figures of merit discussed in this review for each of the technologies discussed. (b) Wavelengths needed for the applications discussed in this review. Black, atomic (dotted) and ionic (solid) lines for atomic species commonly used for quantum information processing. Red, range of wavelengths relevant for biological applications. Blue, range of wavelengths relevant for AR/VR applications.}
\label{Fig:Applications}
\end{center}
\end{figure}
All three of the applications discussed here rely on active control of visible light but each specific application has different performance requirements. To ground our discussion of the current state-of-the-art in visible light modulation, we focus on five metrics of interest: (1) modulation speed; (2) control power per channel; (3) size per channel; (4) optical bandwidth; and (5) technological maturity. Fig.~\ref{Fig:Applications}(a) shows a qualitative description of the importance of each of these metrics for the three application areas and Fig.~\ref{Fig:Applications}(b) shows the wavelengths of interest for applications of visible light modulation. Ideally one technology would provide high speed modulation with low power across the full UV to visible range in a compact form factor. However, there are often trade-offs between these parameters as will become clear in Sec.~\ref{Sec:Surfaces} and Sec.~\ref{Sec:Integrated}. It is likely that each application will require a bespoke or hybrid solution with optimized system-level performance. 
\section{Dynamic Surfaces and Bulk Modulation} 
\label{Sec:Surfaces}

Here we discuss reflective surfaces and bulk devices which enable full control of the intensity and phase distribution of free-space beams. These devices often can control the spatial distribution of the output light as well as the intensity, phase, and frequency of individual beams. They are currently mainstays in the application areas discussed above and provide excellent modulation metrics, though are bulky and often require high power, and there is ongoing work to improve the optical specifications while reducing the size and power consumption. 

\subsection{Dynamic Reflective Surfaces}
\label{Sec:Mech}
In a dynamic reflective surface, an array of controllable reflective pixels modifies the characteristics of an incoming beam of light. Combined with free-space propagation, optical Fourier transforms, and additional spatial filtering these reflective surfaces can be used to control the full image plane~\cite{Nayar} and are thus generally referred to as spatial light modulators (SLMs). As discussed below, the modulation mechanism can be chosen to optimize for speed, optical power efficiency, and degree of control. These systems are mature and commercially available.  Spatial light modulators have been used in a plethora of applications including direct-write lithography~\cite{owa_2023_directwrite}, optical communication~\cite{Sun2021Dec,Cao2019Jul}, fiber amplifiers~\cite{Chen2023Nov}, and microscopy~\cite{Maurer2011Jan}. Research is ongoing to improve pixel density and speed.

\subsubsection{Mirror Arrays}
Mirrors are a fundamental optical element and can modify the direction and phase of an incident laser beam: the orientation of the mirror controls the position of the reflected beam and the position in the travel direction controls the phase. Actuating the stage on which a mirror is mounted thus modulates the position and phase of the incoming laser light. A simple example is the mirror galvanometer which is used for rapidly scanning a single laser beam or image. These are advanced commercial technologies used across industries including confocal microscopy and laser beam machining and are simple, fairly fast, and reliable. However, they are bulky, imprecise, and usually limited to a single or few channels. Alternatively, the mirror can be mounted on a piezo-acutated stage enabling positioning and pointing with nanometer accuracy. This is an indispensable tool for high-resolution feedback control and optical interferometers, including LIGO~\cite{Srivastava2022Mar}.

The advances of micro-electromechanical systems (MEMS) technology has enabled miniaturization of the actuators and the electronic control. MEMS-controlled mirrors reduce required voltages and increase the modulation bandwidth due to their micron-scale size. With these advances, MEMS-controlled mirrors can approach MHz modulation speeds. Such MEMS mirrors can be integrated into a single-mode fiber for binary on/off control or can be assembled into compact one or two-dimensional arrays to form programmable surfaces with locally reconfigurable reflection properties controlled by applying pixel-by-pixel voltages. See \cite{Song2018Jan} for a full review of both commercial and academic work on micromirror arrays. 

MEMS-mirror-based SLMs can control the amplitude or phase of the reflected light. For amplitude control, the orientation of a MEMS-mirror is rotated in a digital or analog fashion. The digital control selectively directs a micron-scale portion of the incident beam to the target image position or to a beam dump, producing the desired intensity pattern. These amplitude-only SLMs are called digital micromirror devices (DMDs). The DMDs were initially used for digital light projection (DLP) technology as licensed by Texas Instruments (TI), but have widely found applications in complex wavefront shaping for optical microscopy~\cite{Fukano2003Jul, Kim_2014, KIM201435, Kim_2019_qrb}, lithography~\cite{lauria_2009_asmldmd}, trapped-ion control~\cite{Shih2021Apr}, optical tweezer generation~\cite{Stuart2018Feb}, and high-precision atomic potential generation~\cite{Zupancic_2016,Tajik2019Nov}. DMDs provide high temporal (32\,kHz) and spatial (4K UHD) resolution and high-extinction ratio between on/off orientations. However, the modulation is generally inefficient as it relies on removing a portion of the incoming light to change the far-field amplitude distribution. 

Mirror arrays with analog rotation control work in a similar fashion and control the reflected wavefront through a controllable diffraction grating with groove and modulation depth programmed by continuously rotating the composite MEMS mirrors. SLMs with rotation-controlled mirrors are generally called grating light valves (GLVs)~\cite{Trisnadi2004Jan} and are licensed by Silicon Light Machines. Like DMDs, GLVs can be used for video display and often offer higher light utilization efficiency. One-dimensional GLVs have modulation speeds up to hundreds of kHz. 

Another important class of mechanical systems is deformable mirrors (DMs). Thousands of MEMS actuators deform the reflective surfaces introducing a local phase on the reflected wavefront. Unlike the DMDs or GLVs used as amplitude SLMs, DMs modulate the phase of the reflected beam, which is lossless in principle. Thus, the main applications of DMs are in optical imaging to actively compensate  aberrations, for instance for due to biological samples in optical microscopy~\cite{Zhang2023Apr} or distortion due to an inhomogeneous atmosphere~\cite{Thompson2002Dec}. Here, the number of actuators determines the degree of control. The modulation speed is comparable to DMDs but the number of actuators is limited to a few thousands. 

\subsubsection{Liquid Crystal Arrays}
When a phase SLM with high spatial resolution is required, a surface composed of nematic liquid crystal on silicon (LCoS) can be used. This comes at the cost of slower frame rate of tens to hundreds of hertz. The refractive index of the LC medium depends on the orientation of the LC molecules with respect to the propagation direction of light. LC molecules align to an external electric field and thus the orientation can be tuned by applying a voltage across the LC region. As the LC molecules rotate, the refractive index of the LC medium changes, imparting a voltage-dependent phase delay on the reflected beam. These devices have been extensively reviewed previously~\cite{Zhang2014Oct,Lazarev2019May} and we discuss integrated versions of these devices in Sec.~\ref{Sec:IntLC}.

\subsection{Transmissive Modulators}
In a transmissive modulator, a bulk crystal is used to modulate the phase, frequency, and amplitude of a single beam though they can be used to provide spatial control as well. These commercial solutions are often fast, but suffer from high per-channel power and size. 

\subsubsection{Acousto-Optic Modulators}
\label{Sec:AOM}
The acousto-optic modulator (AOM) underpins many applications, including communication, Q-switching in lasers, and fast control of atomic systems for quantum technologies. In an AOM, Brillouin scattering~\cite{Brillouin1922} is used to modulate the position, amplitude, frequency, and phase of an incident laser~\cite{Bass2009Sep}. A piezoelectric transducer generates an acoustic wave in a crystal from an input radiofrequency (rf) signal. This acoustic wave creates a periodic modulation in the refractive index of the crystal, acting as a controllable diffraction grating for an incident laser beam. The diffracted component of the laser inherits the phase, frequency, and amplitude of the rf signal and the displacement angle is determined by the modulation frequency. 

The frequency-dependent positioning can be used to easily separate modulated and un-modulated portions of the beam. An AOM can be used as an imaging device by carefully engineering the rf frequency components~\cite{Liu2023Jan}. Moreover, acousto-optic modulation based on slow shear waves enables deflecting the incident beam into a wide range of angles with hundreds of resolvable spots~\cite{MAAK2023129213}. These acousto-optic deflectors are central for neutral-atom quantum information processors as discussed in Sec.~\ref{Sec:Applications}. 

Current commercial systems boast diffraction efficiencies of 1000:1 and switching speeds of below 10\,ns. However, due to their bulk nature, watts of rf power is needed per channel~\cite{Savage2010Oct} which is unfavorable for independent control of a large number of laser beams.  As discussed further in Sec.~\ref{Sec:IntStrain}, there has been some initial work to integrate these systems to reduce the size and power. Hybrid solutions are discussed in Sec.~\ref{Sec:Outlook}. 

\subsubsection{Electro-Optic Modulators}
Electro-optic (EO) modulation broadly refers to a refractive index change due to an external electric field though the exact physical mechanism varies and depends on the specific material~\cite{Sinatkas2021Jul}. Many materials exhibit electro-optic nonlinearities and fast EO modulation is routine at telecom wavelengths. In particular, Lithium Niobate (\LiN) has long been the workhorse for bulk light modulation. \LiN has a relatively large bandgap (3.8\,eV), promising low-loss propagation down to around 350\,nm, and bulk \LiN boasts a strong Pockels effect ($r_{33} \sim 31$\,pm/V) ~\cite{Weis1985Aug} which can be used for EO modulation. Conventional \LiN modulators use either bulk \LiN or large waveguides with a weak change in the refractive index defined by titanium diffusion or annealed proton exchange in bulk \LiN~\cite{Armenise1988Apr}. Due to the low index contrast and thus confinement, these devices are bulky and require high voltages. Nevertheless they out-compete current Si-based modulators, especially for applications with no power or size constraints. In contrast to AO modulation, EO modulation can be much faster but does not affect the spatial degree of freedom so the modulated and unmodulated light cannot easily be separated. 

\subsection{Summary and Outlook}
\begin{figure}[htbp]
\begin{center}
\vspace{0.5cm}
\includegraphics[width=0.48\textwidth]{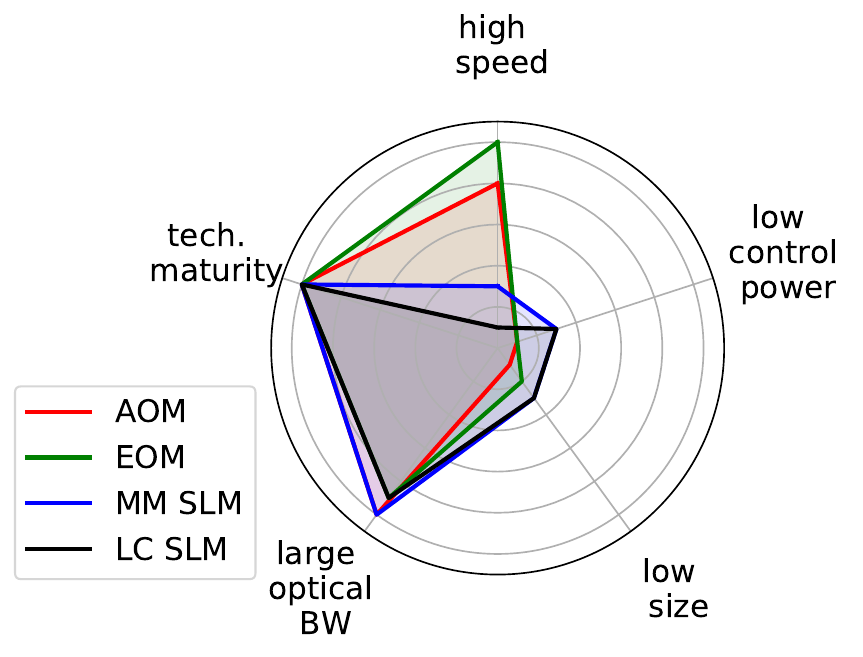}
\caption{\small Qualitative depiction of the capabilities of bulk acousto-optic and electro-optic modulators (AOM, EOM), and liquid crystal and micromirror spatial light modulators (LC SLM, MM SLM) for each of the five figures of merit discussed in this review.}
\label{Fig:DS}
\end{center}
\end{figure}
All the applications discussed in Sec.~\ref{Sec:Applications} currently rely on the dynamic surfaces and bulk modulation technologies discussed in this section. These systems are mature with well-engineered commercial solutions that enable quick integration into larger application systems. However, no technology provides high speed, compact control. AOMs provide high speed single-channel modulation but the large per-channel size and rf power are prohibitive. In contrast, dynamic surfaces provide full image control with a single device but are generally limited to tens of kHz. Fig.~\ref{Fig:DS} shows a qualitative depiction of the four technologies discussed - micromirror spatial light modulators (MM SLMs), liquid crystal spatial light modulators (LC SLMs), and bulk AOMs and EOMs - in the context of the figures of merit of this paper - technological maturity,  modulation speed, electrical control power, size, and optical bandwidth. Fig.~\ref{Fig:IntTotal} puts the optical bandwidth and speed of these commercial technologies in the context of all the technologies discussed in this review. 

While these bulk technologies provide an easy way to modulate the amplitude, phase, and often spatial distribution of light, they clearly lack the integrability in terms of size and power consumption needed for next generation quantum, biological, and augmented-reality applications. Current research is ongoing to provide full spatial-temporal control over light with the necessary speed, and optical bandwidth with reduced electrical power consumption. Optical phase arrays~\cite{Sun2013Jan} are one solution to provide fast, spatial control of light. These rely on visible active integrated photonics which are discussed below in Sec.~\ref{Sec:Integrated}. Active metasurfaces have also been demonstrated for spatial light modulation~\cite{Shaltout2019May}. Finally, resonant surfaces can also improve speed -- a recent work uses photonic crystal cavity pixels, achieving nanosecond pixel switching for a 2D array~\cite{Panuski2022Dec}. As the optical pixel-size is reduced electronic wiring and crosstalk may become the ultimate resolution bottleneck, though this could be overcome with further optical engineering~\cite{Chen2023Aug}.

\section{Integrated Photonics}
\label{Sec:Integrated}

Here we provide an overview of recent advances in active visible photonic integrated circuits (PICs). Visible PICs have applications in low linewidth lasers, optical frequency combs, and sensing. These applications have been covered in a previous review~\cite{blumenthal_2020_review} and are not discussed here. Instead, we focus on devices that provide the amplitude and phase control needed for the applications laid out in Sec.~\ref{Sec:Applications}. In particular, we review four nonlinearities used for integrated visible light modulation - thermo-optic tuning (Sec.~\ref{Sec:IntThermal}), electro-optic tuning (Sec.~\ref{Sec:IntElectro}), piezo-optic tuning (Sec.~\ref{Sec:IntStrain}), and liquid-crystal-based modulators (Sec.~\ref{Sec:IntLC}).

In contrast to the dynamic surfaces and bulk modulators discussed in Sec.~\ref{Sec:Surfaces}, PICs confine light in wavelength-scale structures in overlapping with the active material. This reduces the per-channel size, weight, and power of these active devices and can improve phase stability. However, PICs can suffer from increased optical loss, especially at visible wavelengths. 

In general, optical loss in a PIC is due to a combination of (i) material absorption; (ii) scattering due to device fabrication imperfections; and (iii) scattering due to material defects. Loss due to all three of these sources increase at shorter wavelengths: absorption increases near a material's band edge and Rayleigh scattering due to surface roughness or material imperfections  scales as $\lambda^{-4}$~\cite{Corato-Zanarella2024Feb,West2019Feb}. Nonetheless, low-loss propagation is possible, especially with the field's continual advances in material growth and processing. In this section we will discuss active PIC devices with a guiding layer of Silicon Nitride, Aluminum Nitride, or Lithium Niobate each of which has its own benefits and drawbacks. Other wide bandgap materials such as Aluminum Oxide (Al$_2$O$_3$)~\cite{Hendriks2021Jan}, and diamond~\cite{Schroder2016Apr} can support visible PICs, though we do not discuss them here as there have not been demonstrations of active modulation with these material systems. 

\subsection{Principle of Operation}
\begin{figure}[!bthp]
\begin{center}
\vspace{0.5cm}
\includegraphics[width=0.48\textwidth]{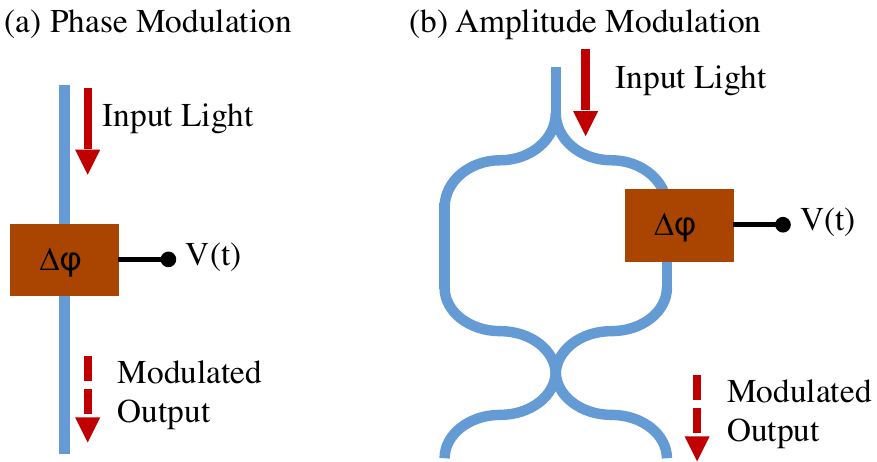}
\caption{\small (a) Method for phase modulation - an external signal changes the index of refraction of portion of a waveguide, adding a controllable phase to the light through the waveguide. (b) The phase modulation creates amplitlude modulation when embedded in a Mach Zender interferometer.}
\label{Fig:IntMod}
\end{center}
\end{figure}

Before discussing particular results, we briefly lay out the basic principle of operation for integrated modulators. In all integrated modulators, an external control is used to modulate the refractive index of a material, which in turn changes the effective refractive index of a guided mode, as illustrated in Fig.~\ref{Fig:IntMod}. This enables control of the phase of the optical mode. Embedding such a device in an interferometer enables amplitude as well as phase control. Device design can optimize parameters of interest such as modulator speed, size, or power consumption. However, in this review we focus on the modulation mechanism and materials that make visible light modulation possible and not on advances in device geometry.



\subsection{Integrated Thermo-Optic Modulators}
\label{Sec:IntThermal}
\begin{figure}[!tbhp]
\begin{center}
\vspace{0.5cm}
\includegraphics[width=0.48\textwidth]{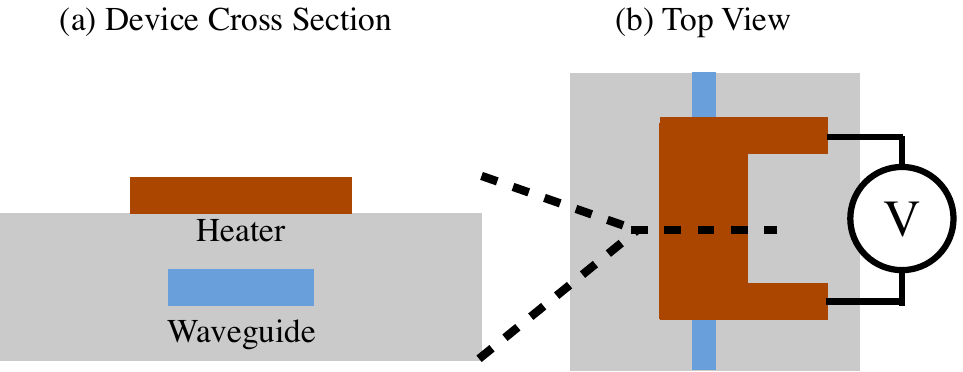}
\caption{\small (a) Simplified cross-sectional diagram of an example electro-optic phase modulator, consisting of a waveguide with control electrodes on each side (not to scale). (b) Simplified top-view schematic of the device.}
\label{Fig:IntTO}
\end{center}
\end{figure}

Thermo-optic (TO) tuning provides a straight forward way to change the real part of a waveguide's effective refractive index, and is compatible with all materials. As seen in Fig.~\ref{Fig:IntTO}, in a TO modulator, a resistive heater changes the temperature, and thus refractive index, of a photonic device. Due to its simplicity and versatility, TO modulation has been extensively employed to provide compact phase and amplitude control. 

TO modulators have been used extensively with Silicon Nitride (SiN$_x$) photonic devices. SiN boasts record-low propagation loss~\cite{Bauters2011Feb} but lacks even-order nonlinearities due to its centrosymmetric molecular structure. Stoichiometric \SiN deposited by low pressure chemical vapor deposition (LPCVD) is transparent above about 400\,nm. High-temperature annealing removes hydrogen and reduces defect absorption. 
\SiN  is also CMOS-compatible and structures can be fabricated on the wafer-scale. 
The vast field of SiN integrated photonics has been reviewed extensively~\cite{Blumenthal2018Sep, Xiang2022Jun, Sharma2020Oct}. Many TO devices are used for static control. Here, we focus on TO modulation results providing switching speeds at or above 1\,kHz.

SiN's fairly large bandgap and mature fabrication techniques have enabled integrated active devices with operating wavelengths below 500\,nm, necessary for all three of the applications discussed in Sec.~\ref{Sec:Applications}. Devices with 50\,kHz switching bandwidth have been demonstrated for applications in neural stimulation~\cite{mohanty_reconfigurable_2020}. Additionally, mature dSiN fabrication enables system-scale fabrication with high channel counts for visible-light optical phased arrays~\cite{chulshin_2020_blue}.  

TO modulators suffer from thermal crosstalk and an innate tradeoff between power consumption and switching bandwidth. Standard devices are hundreds of micrometers in length and require tens of mW for a $\pi$ phase shift. However, thermal and optical engineering can improve these specifications~\cite{alemany_2021_thermal}. For example, power consumption below 1\,mW for $\pi$ phase shift ($W_{\pi}$) has been demonstrated using suspended devices and multi-pass waveguides, though with switching times of only $\sim1$\,kHz~\cite{yong_2022_multipass,miller_2020_multipass}. Similarly, microresonator based TO modulators leverage resonant enhancement to reduce power consumption by a factor proportional to the resonator's finesse. This technology has been used to demonstrate devices with 3dB bandwidths of nearly 300\,kHz with $W_{\pi} \sim 1$\,mW~\cite{liang_2021_vis}. In general, \SiN TO modulators provide a relatively mature solution for applications which can tolerate lower speeds and don't have strict power or temperature constraints.


\subsection{Integrated Electro-Optic Modulators} 
\label{Sec:IntElectro}

\begin{figure}[!tbhp]
\begin{center}
\vspace{0.5cm}
\includegraphics[width=0.48\textwidth]{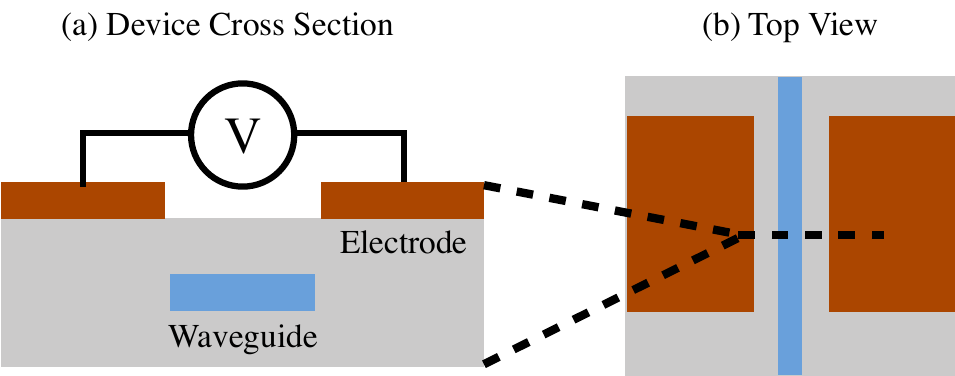}
\caption{\small (a) Simplified cross-sectional diagram of an example electro-optic phase modulator, consisting of a waveguide with control electrodes on each side (not to scale). (b) Simplified top-view schematic of the device.}
\label{Fig:IntEO}
\end{center}
\end{figure}

Here we review integrated versions of the bulk electro-optic modulators discussed in Sec.~\ref{Sec:Surfaces}. Independently of the material used, an electro-optic modulator comprises a waveguide and two electrodes. An common geometry for a phase-only electro-optic modulator is seen in Fig.~\ref{Fig:IntEO}. A voltage applied between the electrodes creates an electric field across the waveguide which creates a local change in the index of refraction of a material with an electro-optic nonlinearity. 

\begin{figure*}[hbtp]
\begin{center}
\vspace{0.5cm}
\includegraphics[width=0.92\textwidth]{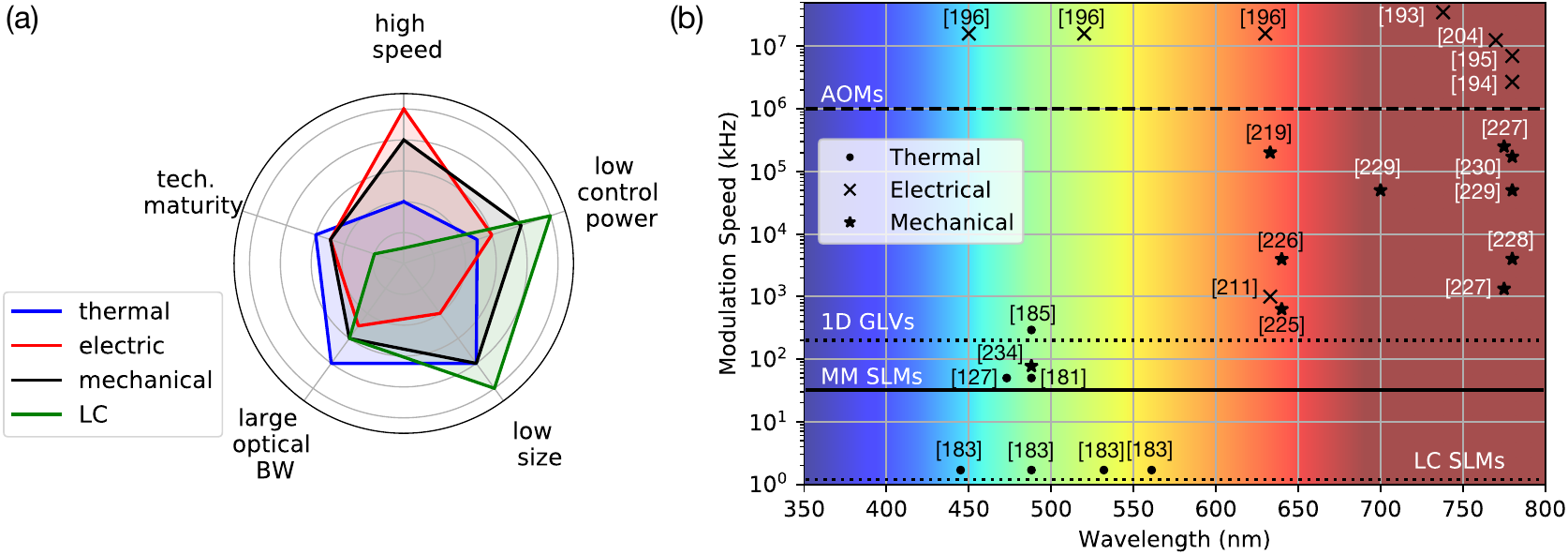}
\caption{\small (a) Qualitative depiction of the capabilities of integrated modulators using thermal, electric, mechanical, and liquid-crystal modulation. (b) Wavelength vs modulation speed for all technologies discussed in this review. Lines represent bulk technologies discussed in Sec.~\ref{Sec:Surfaces}.}
\label{Fig:IntTotal}
\end{center}
\end{figure*}

\subsubsection{Lithium Nioabte}
As discussed in Sec.~\ref{Sec:Surfaces}, bulk \LiN modulators are the workhorse of commercial light modulation but integrated \LiN devices have until recently been elusive due to the difficulty of creating and patterning thin films. However, developments in commercial production of thin film \LiN-on-insulator at the wafer scale~\cite{Luke2020Aug} and fabrication methods~\cite{Siew2018Feb} have spurred a recent surge of integrated photonic devices across the optical spectrum~\cite{zhang_2021_review,Qi2020Jun,Zhu2021Jun,wang_2018_lncmos}. In particular, three recent publications have demonstrated fast modulation in the near-infrared. In \cite{desiatov_2019_visible}, a rib waveguide is used to reduce loss due to sidewall roughness and defects at the \LiN-oxide cladding interface while retaining single mode propagation at 635\,nm and low bending loss. With these devices they demonstrated high speed switching at 850\,nm with a 3\,dB cutoff at 10\,GHz and a $V_\pi=8$\,V for a 2\,mm device (1.6\,V-cm). Further device improvements enabled sub-V-cm, 35\,GHz bandwidth switching at 780\,nm~\cite{Renaud2023Mar}. In \cite{celik_2022_visible} Mg-doped \LiN on a sapphire substrate is used to reduce rf losses, and device design and improved fabrication enabled reduced optical losses, device size, and voltage requirements. These improvements enable modulation of 780\,nm light at 2.7\,GHz and the CMOS-compatible voltage of 4.2\,V for a 3\,mm device (1.4\,V-cm). Finally, \cite{Christen2022Aug} uses full system-level optimization to demonstrate a multi-channel architecture with modulation of 780\,nm at 7\,GHz with 2.2\,V control voltage for a 3\,mm device (0.66\,V-cm) and larger than 20\,dB extinction. Recently, full-spectrum visible modulation was demonstrated at 450, 520, and 630\,nm with sub V-cm control voltages~\cite{Xue2023Jan}.

All of these results boast fairly low switching voltages and large switching bandwidths. This comes at the expense of fairly large device size (mm-scale).  Due to the large bandgap of \LiN, there is no fundamental reason that these results cannot be extended to lower wavelengths although the fabrication will be more difficult and etch roughness will lead to more loss.

Unfortunately, \LiN integrated photonics are further plagued by phase drifts, presumably due to the photo-refractive effect~\cite{Jiang2017Sep,Christen2022Aug, Xu2021Feb} in which the guided optical fields themselves change the local refractive index. This is presumably due to fluctuations of trapped charges, although the exact mechanism is still unknown.  This causes slow drifts in the operating parameters of modulators, especially at high optical powers~\cite{Wang2021Nov,Gozzard2020Jul}. Work has been done to mitigate this effect by removing the cladding~\cite{Xu2021Feb} or through active stabilization~\cite{Christen2022Aug}. Additionally, the pyro-electric coefficient in \LiN can limit fabrication  and is especially a concern for cryogenic integrated photonics~\cite{Bravina2004Jan,Herzog2008Feb} as it can unpredictably change the electrical and optical properties during the cool down from room temperature~\cite{Thiele2023Jun}. Although large-scale systems have been demonstrated despite these issues~\cite{Christen2022Aug}, more research is necessary to understand and mitigate these effects. 

\subsubsection{Aluminum Nitride}
Aluminum nitride (AlN) also offers a relatively high electro-optic coefficient of $\sim$1\,pm/V~\cite{Xiong2012Jul}. High-quality films can be deposited at the wafer scale and AlN deposition and etching is CMOS-compatible. AlN has an extremely wide direct bandgap of 6.12\,eV~\cite{Li2003Dec} with potential to provide low-loss propagation into the UV~\cite{Taniyasu2006May}. In practice, however, scattering losses dominate and waveguides generally exhibit significantly higher loss than in SiN devices. The current record-low loss (still 75\,dB/cm) below 400\,nm is obtained with AlN grown by plasma vapor deposition on a sapphire substrate~\cite{Lu2018Apr}. \cite{Liu2023Mar} provides an in-depth review of AlN photonic devices. AlN modulators have been demonstrated in the infrared, and one study reported modulation of 770\,nm light with a modulation bandwidth of 12.5\,GHz with a microring resonator limited by the detection, not device, bandwidth~\cite{Xiong2012Jul}. However, improved material quality and fabrication processes are needed to reduce propagation losses extend the optical bandwidth of these devices to their bandgap limit. 

\subsubsection{Emerging Materials}

We have focused on \LiN and AlN due to their prevalence as integrated optics platforms, but other materials are emerging as possible solutions for fast EO modulation in the visible spectrum. For instance, Barium Titanate (BTO) has a large bandgap of 3.2\,eV and a strong EO coefficient of up to 1300\,pm/V for unclamped films~\cite{Karvounis2020Dec,Abel2019Jan}. BTO films deposited on Magnesium Oxide substrates have been shown to have low loss at 633\,nm and large ridge-waveguide EO modulators were fabricated providing up to 1\,MHz modulation at 633\,nm~\cite{petraru_2003_mgobto}. Graphene can also be used for EO modulation with incredibly compact form factors (10s of $\mu$m) and fast modulation (10s of GHz) for single devices~\cite{phare_2015_graphene,Liu2011Jun}, and even on the wafer scale~\cite{giambra_2021_wafergraphene}. Although these results focused on infrared modulation, similar techniques could be translated to modulation of visible light as required for the applications discussed in this review. Finally, Silicon Carbide (SiC), a wide bandgap material, is emerging as a platform for integrated photonics~\cite{Yi2022Sep}, especially as it is CMOS-compatible. Recent improvements in fabrication quality enabled GHz modulation at infrared wavelengths in 3C-SiC~\cite{Powell2022Apr} and further fabrication improvements could enable visible modulation as well.

\subsection{Integrated Strain-Tuning-Based Modulators} 
\label{Sec:IntStrain}

In this section we discuss devices which use physical deformation of photonic structures for visible light modulation. This effect takes many forms depending on the material platform. In all instances a change in the local refractive index or device geometry due to strain is used to modulate light routed through a waveguide. 

As discussed in Sec.~\ref{Sec:AOM}, AOMs, which make use of Brillouin scattering to modulate light, are a commercial mainstay for visible light modulation. While these commercial products are generally based on bulk crystals, the first demonstrations of acoustic-optic devices focused on modulation of light in a 2D guided mode with surface acoustic waves. In a seminal publication in 1970, 633\,nm light was guided in a thin glass film deposited on the surface of a $\alpha$-quartz crystal providing modulation efficiencies of up to 66\% with 0.18\,W of acoustic power~\cite{Kuhn1970Sep}. In 1974, a device employing 2 tilted surface acoustic waves in an optical guiding layer of \LiN enabled high bandwidth (200\,MHz) modulation of 633\,nm light with a  diffraction efficiency of 50\% at 0.2\,W rf power~\cite{Tsai1975Feb} and later device improvements enabled amplitude and frequency modulation without position modulation, imperative for applications which call for fiber and waveguide coupling~\cite{Cheng1992Jan}. Because these guided-wave acoustic-optic devices confine both optical and acoustic waves in a 2D planar structures, the light-sound interaction is enhanced, improving efficiency, bandwidth, and power consumption.

More recently, an integrated AOM built on \LiN was demonstrated in the infrared for LiDAR systems~\cite{Li2023Aug}. In this demonstration, the acoustic wave and optical wave are anti-parallel -- in contrast to most commercial devices and previous planar devices where the acoustic wave propagates perpendicularly to the optical field. With the anti-parallel geometry, the light is directly diffracted out of the chip. In this device, the authors achieve a switching bandwidth of 0.7\,MHz, likely limited by electronics. This technology could conceivably be translated to the visible, with direct applications in atom control~\cite{Lin2023May} which, as discussed in Sec.~\ref{Sec:Applications} relies on 2D beam positioning to move the atoms for different operations. Patterning waveguides to achieve full confinement could further improve optical performance. In a recent demonstration, on-chip acoustic modulation was used to frequency shift and deflect an integrated optical field, an important step towards a fully integrated device~\cite{Shao2020Aug}. Again, although demonstrated at 1597\,nm, the \LiN platform should allow for direct translation to visible modulation.

In contrast to the above works which use an RF field to induce a controllable grating in a piezo-optic material, more recent works directly deform a nano-photonic device. This idea was initially discussed in \cite{vanderSlot2019Jan} which considered modulation of a buried SiN waveguide. 

In \cite{tian_2020_sinstiff}, the authors use high-overtone bulk acoustic wave resonances (HBARs) to tune a buried SiN ring modulator. Critically, the acoustic mode in an HBAR is laterally confined, unlike the surface acoustic waves used in the devices discussed previously, enabling compact devices with low crosstalk. This device achieves a 1\,GHz bandwidth at 21\,dBm power consumption. While this device operates at 1550\,nm, the widebandgap of SiN would allow direct translation to visible wavelengths.

In 2015, the commercial SiN-based Triplex process from Lionix was extended to include lead zirconate titane (PZT) to modulate the stress in the SiN waveguide structure. With this, they demonstrated modulators at 640\,nm with 4\,MHz modulation bandwidth~\cite{Hosseini2015Jun}.  While fast and relatively low power (0.3\,$\mu$W at near-DC modulation), these devices are large at 8-10\,mm each. A similar system of SiN integrated with PZT modulation was used to demonstrate modulation bandwidths above 33\,GHz in the infrared spectrum~\cite{Alexander2018Aug}.

A recent set of publications demonstrated integrated modulation of SiN devices at visible wavelengths using AlN as the piezoelectric material. In \cite{stanfield_2019_peizosin}, a 40\,$\mu$m diameter SiN ring resonator is modulated at 250\,MHz with 0.125\,mW of electrical power at room temperature, and even lower power consumption when operated at 7\,K. Additionally, a waveguide-based Mach-Zender modulator (MZM) is modulated in 750\,ns with 20\,nW power dissipation. However, these MZMs suffered from large insertion loss and high voltage requirements, leading to a 175\,VdB voltage-loss product and limiting the utility in large scale systems. In a follow-up work, an undercut of the modulated structures improved the voltage requirements and improved loss to achieve a 0.8\,VdB voltage-loss product~\cite{dong_2022_piezosin2}, although with a modulation speed limited to 4\,MHz by the mechanical resonance. In \cite{dong_2022_piezosin} simultaneous optimization of all optical, electrical, and mechanical components results in devices with 50\,MHz modulation with a power dissipation of only 200\,$\mu$W at 1\,MHz modulation. Finally, resonant modulators provide a smaller footprint with a larger modulation bandwidth (172\,MHz) and high extinction ratio ($>$\,30\,dB)~\cite{Menssen2023Oct}. These devices are the first designed and tested with the requirements for atomic quantum control, considering repeatability and scalability as well as device-specific parameters.


Similarly, micro-electromechanical systems (MEMS) can be used to physically move nanophotonic structures, inducing modulation. Silicon MEMS can change the distance between waveguides in a directional coupler, enabling repeatable amplitude modulation at MHz speeds and impressive scalability~\cite{Seok2019Apr} with applications in LIDAR~\cite{Zhang2022Mar} or post-fabrication cavity tuning~\cite{Mouradian2017Apr}. A recent experimental paper demonstrated active positioning of visible light using a SiN guiding layer on MEMs cantilever devices~\cite{SharifAzadeh2023May}. These structures achieved 1D and 2D positioning of 488\,nm light at 78\,kHz.

\subsection{Integrated Liquid-Crystal-Based Modulators}
\label{Sec:IntLC}

Compared to integrated thermo-optic, electro-optic, and piezo-optomechanical modulators, integrated liquid-crystal-based (LC-based) modulators enable very compact form factors and power-efficient operation, at the expense of slower modulation speeds (typically sub-kHz)~\cite{atsumi19oe, xing15ptl}. These LC-based modulators leverage the large birefringence and electrically-tunable refractive index of LC media by integrating LC into integrated-photonics platforms. 

\begin{figure}[!bthp]
\begin{center}
\vspace{0.5cm}
\includegraphics[width=0.48\textwidth]{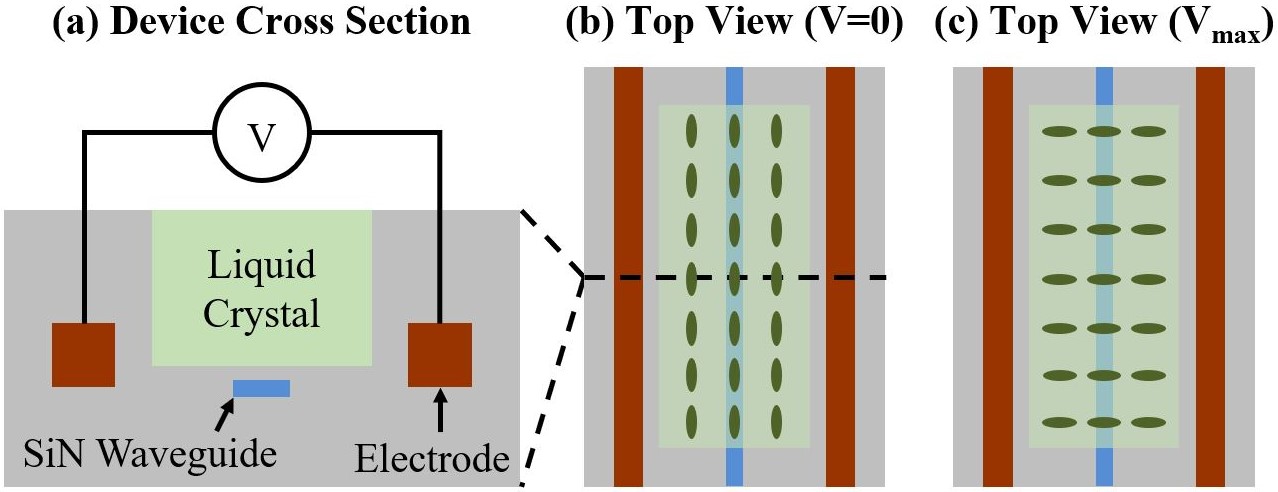}
\caption{\small (a) Simplified cross-sectional diagram of an example integrated LC-based phase modulator, consisting of a silicon-nitride waveguide recessed within silicon dioxide below an LC-filled trench with integrated electrodes on each side (not to scale). (b) Simplified top-view schematic of the device when no voltage is applied across the liquid-crystal region, resulting in parallel alignment of the LC molecules with respect to the waveguide (not to scale). (c) Simplified top-view schematic of the device when the maximum voltage is applied across the LC region, resulting in complete rotation of the LC molecules and perpendicular alignment of the molecules with respect to the waveguide (not
to scale). \cite{notaros22oe}}
\label{Fig:IntLC}
\end{center}
\end{figure}

As shown in Fig.~\ref{Fig:IntLC}, these LC-based modulators typically consist of a waveguide that guides the light beneath a trench filled with LC medium, which enables the optical mode of the waveguide to overlap and interact with the integrated LC medium. Consequently, the effective refractive index and profile of the optical mode are influenced by the refractive index of the LC region. The refractive index of the LC medium seen by the optical mode depends on the orientation of the LC molecules with respect to the direction of light propagating in the waveguide. The orientation of the LC molecules can be tuned by applying a voltage across the LC region using integrated electrodes, since LC molecules align to an external electric field (as discussed above in Sec.~\ref{Sec:IntLC}). As the LC molecules rotate, the refractive index of the LC medium changes, resulting in a change in the effective index and profile of the optical mode.

Various commercially-available LC compounds can be used for these modulators, with the compound used impacting parameters such as the birefringence at a given wavelength, the operating temperature range, and the modulation speed. For example, a commonly-utilized LC compound is 5CB, which has ordinary and extraordinary refractive indices of 1.53 and 1.7, respectively, at red wavelengths and a nematic-state operating temperature range of 18 to 35$^\circ$C.

Integrated LC-based demonstrations have largely focused on operation at infrared wavelengths~\cite{atsumi19oe, xing15ptl, pfeifle12oe, cort11ol, ptasinski13ol, maune03apl, iseghem20ecio, falco07oc, wang13oe, dai15oc}. Recently, integrated LC-based modulators have been shown at visible wavelengths by integrating nematic LC media into silicon-nitride-based platforms~\cite{notaros22oe, notaros18fio, notaros24ol, notaros19ipr, coleto23ipc}, enabling compact phase and amplitude modulation at a red operating wavelength. For example, an integrated LC-based phase modulator achieved $41\pi$ phase shift within $\pm2.4$ V for a 500-$\upmu$m-long modulator, which means that a $2\pi$ phase shifter would need to be only 24.4 $\upmu$m long (corresponding to a 0.003 V-cm modulation requirement)~\cite{notaros22oe, notaros18fio}. Moreover, an integrated LC-based amplitude modulator, which leverages the LC birefringence to tune the coupling coefficient between two waveguides, achieved amplitude modulation with a device length of only 14 $\upmu$m~\cite{notaros19ipr,notaros24ol}.

Due to the large birefringence afforded by LC media, these integrated LC-based devices are one to two orders of magnitude more compact than comparable thermo-optic, electro-optic, or piezo-optomechanical modulators at visible wavelengths. While longer modulators may be adequate for certain systems and applications, many systems require compact modulators, such as integrated optical phased arrays that require dense integration for high fill factors. As such, these compact and power-efficient integrated LC-based modulators are well suited to enable densely-integrated visible-light systems for many impactful applications areas~\cite{corsetti24lsa, corsetti22ipr, notaros23ol, notaros19cleoPassive, desantis23ipc}, including augmented-reality displays (as discussed in Sec.~\ref{Sec:Applications}). However, for applications requiring high-speed modulation, including most quantum and biological applications, the modulation speed of these integrated LC-based modulators will need to be improved. 

\subsection{Summary}
In contrast to the devices discussed in Sec.~\ref{Sec:Surfaces}, the integrated photonic systems discussed in this section are less mature, but promise scalability due to significant improvements in size and per-channel power. Fig.~\ref{Fig:IntTotal}(a) presents a qualitative overview of the four main integrated technologies discussed in this review. There is a clear trade-off between different figures of merit, most obviously between the control power and speed of modulation. 

Similarly, Fig.~\ref{Fig:IntTotal}(b) presents an overview of all technologies and publications discussed in this review with respect to operating wavelength and the modulation speed. While there has been significant work in developing visible light integrated photonics, there is a clear gap in demonstrating fast, low-wavelength modulation, particularly critical for quantum applications. Moreover, most published works focus on single-device characteristics and systems-scale optimization will be needed before they become ubiquitous in application spaces.

\section{Outlook and Conclusion} 
\label{Sec:Outlook}
As outlined in this review, there have been remarkable advances in active integrated photonics over the last decade. However, Fig.~\ref{Fig:IntTotal} clearly demonstrates that there is significant work to be done before these platforms are ready for deployment into quantum, biological, and commercial augmented-reality display systems. In this final section, we outline some important next steps if active visible photonics are to reach the technology readiness level needed for widespread adoption. 

Throughout this review, we have focused on single-modality solutions. However, as is evident in Fig.~\ref{Fig:DS} and Fig.~\ref{Fig:IntTotal}, no one technology provides the needed performance across the figures of merit highlighted here: speed, power, optical bandwidth, size, and technological maturity. Indeed, it is likely that a tradeoff will persist even as these technologies mature. Thus, it is likely that hybrid systems will be needed. 

Indeed, hybrid solutions are already being employed to provide simultaneously high temporal and spatial bandwidth. For many neutral-atom quantum computing, biological control, and augmented-reality display applications, the spatial bandwidth, or the number of individually controllable channels, is also important. There is currently no technology enabling both high-speed modulation and high-resolution spatial modulation. AOMs, EOMs, and integrated photonics allow for sub-microsecond modulation but are limited to a few channels. The SLMs based on MEMS or LCoS dynamic surfaces enable light modulation in the spatial domain with millions of resolvable spots, but their modulation speed is limited to tens of kilohertz. 

High-speed single or few-channel modulators integrated with high-resolution SLMs offer an intermediate solution. Specifically, an angle- (or position-) multiplexed hologram displayed on a high-resolution SLM can be illuminated by different spatial modes whose amplitude and phase are modulated by an array of high-speed optical modulators~\cite{Kim_2021_patent}. These intermediate solutions are used for high-speed optical mode sorting~\cite{Braverman_2020}, neutral-atom quantum information processing~\cite{Kim_2021_patent}, and the individual beam shaping of modulated beam arrays~\cite{Christen2022Aug}.

Finally, the results presented in Sec.~\ref{Sec:Integrated} have, for the most part, been device-scale demonstrations. However, all of the applications discussed in Sec.~\ref{Sec:Applications} necessitate multiple channels and full system packaging. This includes coupling light out of the chip with high efficiency, simultaneous control of multiple channels with low crosstalk, and integrated driving electronics. Additionally, while we focused on five figures of merit in this review, it will also be important to improve other capabilities such as on-off isolation, optical power handling, and device crosstalk. As these technologies mature from device-scale demonstrations to adoption by the applications community it will be necessary to optimize system-scale specifications. 

To summarize, active visible-light modulation is an enabling technology for a wide variety of fields. Here we focus on quantum, biological, and augmented-reality display applications, but it is likely that, as control of visible light matures, the application space will increase in step. There are a number of promising platforms for high-speed, low-power visible-light control and we anticipate that the research in novel methods to modulate visible light will spur innovations in application spaces and vice versa.  \\

\noindent\textbf{Acknowledgements}:
We are grateful for helpful discussions with Ian Christen while preparing this review. S.M. acknowledges support from the National Science Foundation (NSF) Grant No. ECCS-2240291. A.M. acknowledges support from AFOSR. This material is based upon work supported by the Air Force Office of Scientific Research under award number FA9550-23-1-0245. S.P. and D.K. acknowledge support from the National Research Foundation of Korea (NRF) grant funded by the Korea government (MSIT)(NRF-2022M3E4A1074937). J.N. and M.N. acknowledge support from the National Science Foundation (NSF) Faculty Early Career Development (CAREER) Program (Grant No. 2239525).

\bibliography{main}
\end{document}